\documentclass{emulateapj}
\usepackage{natbib}
\usepackage{times}
\usepackage[english]{babel}

\slugcomment{{Submitted to ApJ.}}

\usepackage{epsfig}
\usepackage{amsmath}
\usepackage{amssymb}
\usepackage{graphicx}

\usepackage{savesym}
\usepackage{SIunits}
\usepackage{verbatim}
\usepackage{hyperref}

\restoresymbol{SI}{square}

\def\Gadget2{\rm{\textsc{Gadget\thinspace 2}\ }}
\def\kms{{\rm\thinspace km\thinspace s}^{-1}}
\def\eV{{\rm\thinspace eV}}

\def\kpc{{\rm\thinspace kpc}}

\def\Msun{\hbox{$\mathrm{\thinspace M_{\odot}}$}}
\def\pc{{\rm\thinspace pc}}

\def\s{{\rm\thinspace s}}
\def\yr{{\rm\thinspace yr}}
\def\Myr{{\rm\thinspace Myr}}

\newcommand{\cm}{\centi\meter}

\newcommand{\muG}{\mu G}
\newcommand{\yt}{{\texttt{yt}}}

\newcommand{\cmsps}{\centi\meter\squared\per\second}

\shorttitle{Cosmic ray driven outflows}
\shortauthors{Hanasz, et al.}

\received{August, 19, 2013}
\accepted{October 11, 2013}

\begin{document}

\title{Cosmic rays can drive strong outflows from gas-rich high-redshift disk galaxies}

\author{M. Hanasz\altaffilmark{1},
           H. Lesch\altaffilmark{2},
           T. Naab\altaffilmark{3},
           A. Gawryszczak\altaffilmark{4,5},
           K. Kowalik\altaffilmark{1},
           D. W\'olta\'nski\altaffilmark{1}}

  \altaffiltext{1}{Centre for Astronomy, Nicolaus Copernicus University,
  Faculty of Physics, Astronomy and Informatics,
  Grudziadzka 5, PL-87100 Toru{\'n}, Poland, mhanasz@astri.uni.torun.pl}
\altaffiltext{2}{Universit\"ats-Sternwarte M\"unchen, Scheinerstr.\ 1, D-81679}
\altaffiltext{3}{Max-Planck-Institut f\"ur Astrophysik,
  Karl-Schwarzschild-Str. 1, D-85741 Garching bei M\"unchen, Germany}
\altaffiltext{4}{Pozna\'n Supercomputing and Networking Centre, ul. Noskowskiego 10, PL-61-704 Pozna{\'n}, Poland}
\altaffiltext{5}{Nicolaus Copernicus Astronomical Center, Bartycka 18, Warsaw PL-00-716, Poland}

\begin{abstract}
\label{abstract}
We present simulations of the magnetized interstellar medium (ISM) in models of
massive star forming ($40 \Msun\yr^{-1}$) disk galaxies with high gas
surface densities ($\Sigma_{\mathrm{gas}} \sim 100 \Msun \pc^{-2}$) similar to
observed star forming high-redshift disks. 
We assume that type II supernovae deposit 10 per cent of their
energy into the ISM as cosmic rays and neglect the additional
deposition of thermal energy or momentum. With a typical
Galactic diffusion coefficient for CRs ($3\cdot 10^{28}\cm^2 \s^{-1}$) we
demonstrate that this process alone can trigger the local formation of
a strong low density galactic wind maintaining vertically open field
lines. Driven by the additional pressure gradient of the relativistic
fluid the wind speed can exceed $10^3 \kms$, much higher than the
escape velocity of the galaxy. The global mass loading, i.e. the
ratio of the gas mass leaving the galactic disk in a wind to the star
formation rate becomes of order unity once the system has settled into
an equilibrium. We conclude that relativistic particles accelerated in
supernova remnants alone provide a natural and efficient mechanism to
trigger winds similar to observed mass-loaded galactic winds in
high-redshift galaxies. These winds also help explaining the 
low efficiencies for the conversion of gas into stars in galaxies as
well as the early enrichment of the intergalactic medium with
metals. This mechanism can be at least of similar importance than the
traditionally considered momentum feedback from massive stars and
thermal and kinetic feedback from supernova explosions.
\end{abstract}


\section{Introduction}
\label{Intro}

The universal rate for the conversion of gas into stars in galaxies peaks at
redshifts z $\sim 1.5 - 3$ (e.g. \citealp{2006ApJ...651..142H}). At this
epoch star forming galaxies drive powerful galactic winds which can transport a
significant fraction of the gas away from the central galaxy making it
temporarily unavailable for star formation
(e.g. \citealp{2000ApJ...528...96P,2003ApJ...588...65S,2009ApJ...692..187W,2010ApJ...717..289S,2011ApJ...733..101G, 
2012ApJ...752..111N, 2012ApJ...761...43N}). Spatially resolved high-redshift
observations indicate that these winds are launched directly from the sites of
the - typically strongly clustered - star formation
\citep{2011ApJ...733..101G,2012ApJ...752..111N, 2012ApJ...761...43N}. The
estimated outflow rates $\dot{M}_{\mathrm{out}}$ can be several times higher than
the star formation rates (SFR). The resulting high mass loading $\eta =
\dot{M}_{\mathrm{out}}/\textrm{SFR}$ indicates that - even at the peak epoch of
cosmic star formation - the amount of gas expelled from the galaxies is
comparable to the amount of gas converted into stars inside the galaxies.

The direct observational evidence for inefficient conversion of gas into stars
is supported by indirect constraints from halo abundance matching techniques.
Here the galaxy formation efficiency can be defined as the fraction of the
stellar mass of a galaxy to the total available baryonic mass of its host dark
matter halo. In a concordance $\Lambda$-CDM cosmology this efficiency peaks -
almost independent of redshift - for galaxies in halos of about $10^{12}\Msun$
and never exceeds $\sim 20 - 25 \%$ (e.g. \citealp{ 2010ApJ...710..903M,2010ApJ...717..379B,
2010MNRAS.404.1111G, 2013MNRAS.428.3121M}).  Therefore at least $3/4$
(significantly more in halos of higher as well as lower mass than
$10^{12}\Msun$) of the baryonic material is never converted into
stars, eventually due to powerful galactic winds. It is plausible that
the main physical processes responsible for driving the outflows also
regulate the efficiency with which the available gas is converted into
the stellar components of galaxies in the Universe.

In connection to the evolution of stellar populations a number of physical
processes are - in principle - energetic enough to expel gas from star forming
galactic disks. Besides AGN for high mass galaxies
\citep{2006MNRAS.365...11C}, type II supernovae have long been
considered the most promising candidates, in particular for lower mass
galaxies \citep{1974MNRAS.169..229L,1986ApJ...303...39D}. Although the amount of energy
per event is significant the thermal energy is mainly deposited at the sites of
star formation, i.e. dense molecular clouds. Here the cooling times are very
short and the energy can be efficiently radiated away making it difficult but
not impossible to drive large scale galactic winds (for recent discussions see
e.g. \citealp{2011MNRAS.415.1051B,2012MNRAS.426..140D}). However, even before
the supernova explosions, the momentum and energy input from massive stars in
the form of stellar winds and stellar luminosity is significant and might support
the wind driving \citep{2005ApJ...618..569M,2012MNRAS.421.3522H,2013ApJ...770...25A}. 

In this paper we focus on a separate mechanism: the formation of large scale
magnetized galactic winds driven by cosmic rays. As supernovae drive strong
shocks into the interstellar medium some fraction of the explosion energy is
consumed to accelerate ionized particles to relativistic energies which are then
injected into the ISM as cosmic rays (CR)
\citep{1977DoSSR.234.1306K,1978MNRAS.182..147B,1978ApJ...221L..29B}. This
relativistic fluid is coupled to the galactic magnetic field and - in particular
the hadronic component - is less prone to energy losses than the gaseous
component of the ISM. Analytic estimates and numerical experiments without or
only approximate inclusions of galactic magnetic fields clearly indicate that
CRs can help driving large scale galactic winds
\citep{1991A&A...245...79B,2002A&A...385..216B,2007ARNPS..57..285S,2008ApJ...674..258E,2012MNRAS.423.2374U,2012A&A...540A..77D}. However, CRs are
strongly coupled to magnetic fields whose evolution should be followed
in a self-consistent
way. \citet{2004ApJ...605L..33H,2009ApJ...706L.155H,2010A&A...510A..97S,2011ApJ...733L..18K} have shown that  
CRs promote buoyancy effects in the interstellar medium, leading to
the break-out of magnetic fields from galactic disks  \citep{1992ApJ...401..137P}
and, at the same time, to magnetic field amplification
by CR-driven dynamo action. Plausibly, such processes are
also relevant for star forming galaxies at high redshift which are observed to
have significant magnetic fields at the level of tens of~$\muG$.
\citep{2008Natur.454..302B}. Recent observations even demonstrate the existence of
large magnetic fields up 50~kpc away from the galaxy indicating strong
large-scale magnetized winds \citep{2013arXiv1307.2250B}. In this
letter we present a three-dimensional full MHD simulation of a massive gas-rich disk
galaxy (section \ref{numerical_setup}) and follow the formation of large scale
magnetized winds as a dynamic response to the injection of CRs (section
\ref{simulations}). This is considered as a proof of principle for the
importance of this physical process. The most important implications
are presented in section \ref{discussion}. 

\section{Numerical setup}\label{numerical_setup}

For the simulations we use the PIERNIK MHD code \citep{piernik1}, a
grid-MHD code based on the Relaxing TVD (RTVD) scheme by
 \cite*{jin-xin-95} and  \cite*{2003ApJS..149..447P}. The induction
equation, including the Ohmic resistivity term, is integrated with a
constraint transport (CT) algorithm \citep{1988ApJ...332..659E}. The
original scheme is extended to include dynamically independent, but
interacting fluids: thermal gas and a diffusive CR gas, described
within the fluid approximation. \citep{2003A&A...412..331H}. We incorporate
selfgravity of interstellar gas and gravitational potential is
obtained by solving the Poisson equation inside the computational
domain with an iterative, multi-grid solver
\citep{doi:10.1137/S1064827598346235} combined with a multipole solver 
\citep{1977JCoPh..25...71J} to properly treat the gravitational
potential at 'isolated' boundaries.

We assume a fixed gravitational field due to the stellar disk and the
dark matter halo and compute the gravitational potential using the model of
\cite{1991RMxAA..22..255A} with $M_{\textrm{halo}} =
8 \cdot 10^{11} \Msun$ within $R_{\textrm{cutoff}} = 100\kpc$ and
$M_{\textrm{disk}} =8.6 \cdot 10^{10} \Msun$. We neglect the contribution of a
central bulge. Fresh gas is supplied to the disk at a fixed
rate  of $\dot{M}_{\textrm{in}} = 100 \Msun \yr^{-1}$  following the
initial gas density distribution.  
To simplify the setup the gas is added directly at the disk plane.  This is
clearly a simplified model.  However, another simplified way of treating the
gas supply by spherical accretion is much more difficult to control. Even in
this case we expect that the highly collimated winds forming in our simulation
would punch through spherically accreting gas not changing our conclusions. The
more realistic alternative of gas accretion along filaments would provide fresh
gas, carrying high angular momentum, at disk peripheries and is not expected to
change the wind properties significantly.

We construct a three-dimensional gas distribution
$\rho_0(x,y,z)$ (e.g. \citealp{1998ApJ...497..759F}). At every time
step we add gas at the given inflow rate $\Delta \rho (x,y,z) =
\dot{M}_{\textrm{in}}/M_{\textrm{0}} \rho_0(x,y,z) \Delta t$ within
the disk volume. The disk collects gas until it becomes
gravitationally unstable. We assume that star formation is controlled
by a star formation efficiency parameter $\epsilon \le 1$. In the
actual models we assume $\epsilon = 0.1$. The star formation rate per
unit volume is computed as 
\begin{equation}
\rho_{\textrm{SFR}} = \epsilon \sqrt{\frac{G\rho^3}{32\pi}}
\end{equation}
provided that gas density exceeds some threshold density
$\rho_{\textrm{thr}}$, which we treat as a free parameter.
Its value ($\simeq 600\textrm{ H atoms }\cm^{-3}$) is adopted
to obtain highly localized star formation, and to regulate the overall
SFR of the galaxy. We locally deplete the gaseous ISM at the same rate at
every timestep.

We assume that one supernova occurs per $100\Msun$ of gas forming new
stars, and that $10\%$ of the explosion energy is used to accelerate
CRs. Individual CR particles propagate at relativistic speeds, however
fast streaming of CRs along magnetic field lines leads to streaming
instabilities \citep{1969ApJ...156..445K}, the generation of
small-scale turbulence, and subsequently the scattering of CRs on 
self-excited turbulence. This implies that a CR pressure gradient term
has to be taken into account in the gas equation of motion.
The bulk motion of CRs is considered as a combination of diffusive
and advective propagation and can be described by the
diffusion-advection transport equation. 
In our model CRs diffuse preferentially along magnetic field lines,
while diffusion perpendicular to the magnetic field is significantly
much less efficient \citep[see][Chap. 9]{1990acr..book.....B}.

A more elaborated picture of CR propagation should include additional effects,
such as energy conversion from CRs to waves, energy sinks for MHD waves due to
ion-neutral collisions and nonlinear Landau damping (for a detailed discussion
see e.g. \citet{2002A&A...385..216B, 2012A&A...540A..77D} and references
therein). The above mentioned  authors adopt  constant CR diffusion
coefficients ranging from $ K= 10^{27}  \cm^2 \s^{-1}$ up to $10^{30} \cm^2
\s^{-1}$  in their 1-D numerical models. The assumption of constant diffusion
coefficient  implies  that the large-scale diffusion velocity $v_\mathrm{diff}
= - K\nabla e_\mathrm{CR}/e_\mathrm{CR} $ may become larger than the Alfv\'en
speed in regions of steep gradients of the CR energy density.  This can occur
in the disk, and especially around CR production regions, where molecular and
neutral gas components dominate.  Since Alfv\'en waves are efficiently
dissipated there by ion-neutral damping, the most important contribution to the
random magnetic field in the disk is induced by supernova explosions.
Therefore ion-neutral collisions should reduce the amplitude of Alfv\'en waves,
leading to enlargement of the mean free path of the CR particles, and therefore
to higher diffusion coefficients.

The numerical algorithm of the anisotropic CR propagation, within the
framework of staggered mesh MHD code, has been described in 
\cite{2003A&A...412..331H}. The values of CR diffusion coefficients,
parallel and perpendicular to the magnetic fields adopted for the actual models are $K_\parallel =
{3\cdot{10}^{28}}{\cmsps}$, $K_\perp= {3\cdot {10}^{26}}{\cmsps}$.
Initially the toroidal magnetic field pervading  the disk has a
strength of $3 \muG$  and a uniform magnetic diffusivity
$\eta={3\cdot{10}^{25}}{\cmsps}$, corresponding to a standard value of
turbulent diffusivity of the ISM.

The simulations have been performed at a resolution of $512^3$ grid cells,
distributed among equal-sized MPI blocks, in the Cartesian domain
spanning a volume of ${100^3}{\kpc^3}$. The disk is placed at the
centre of the domain, and the disk plane is parallel to $x-y$ plane of
the coordinate system. We impose outflow boundary conditions for the
gas component at all domain boundaries. Fixed boundary conditions
($e_{\rm CR}=0$) on external domain boundaries are assumed for the CR
component.

\section{Simulations} \label{simulations}

\begin{figure}[ht]
   \centering
   \includegraphics[width=0.85\columnwidth]{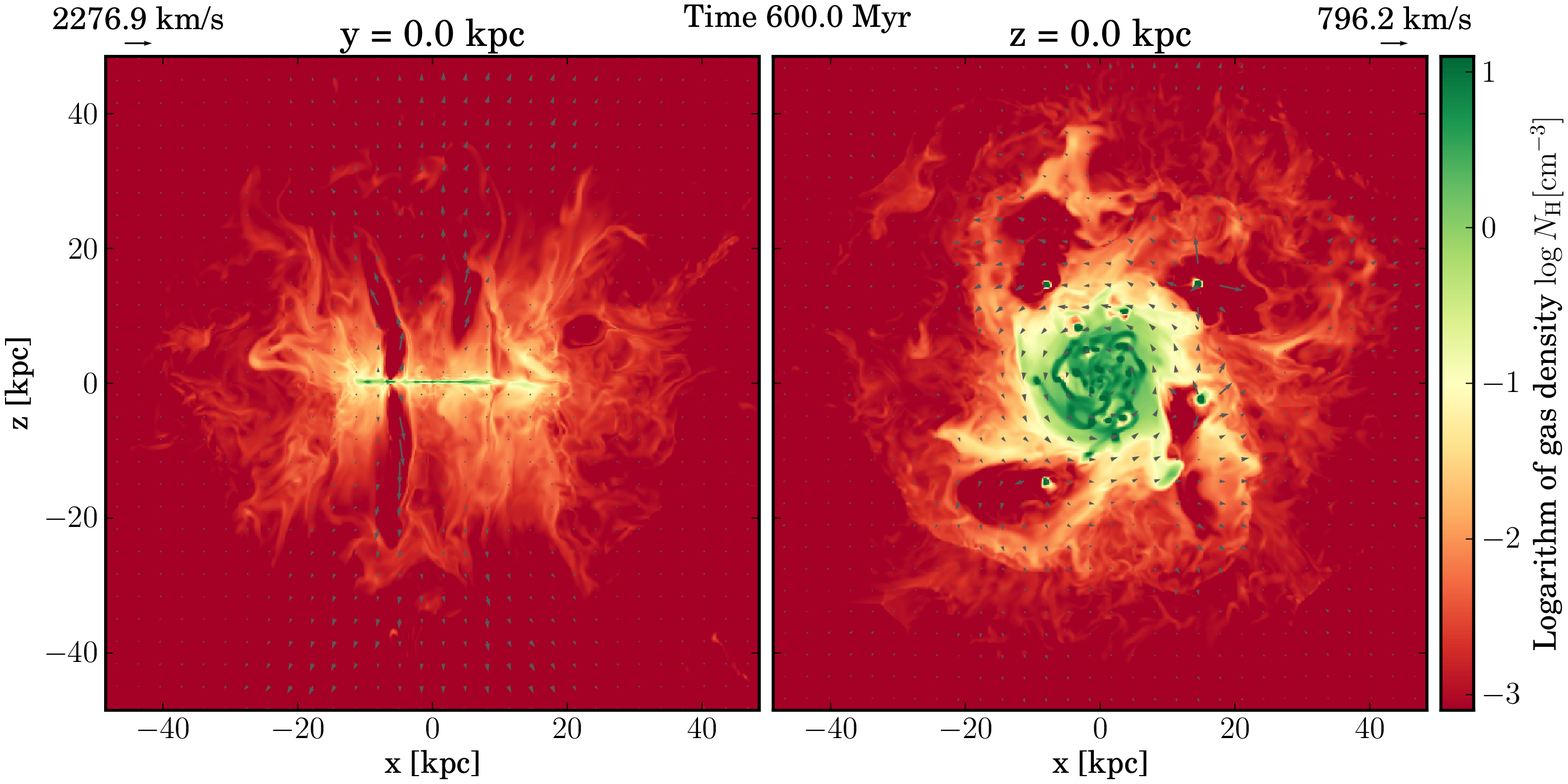}
   \includegraphics[width=0.85\columnwidth]{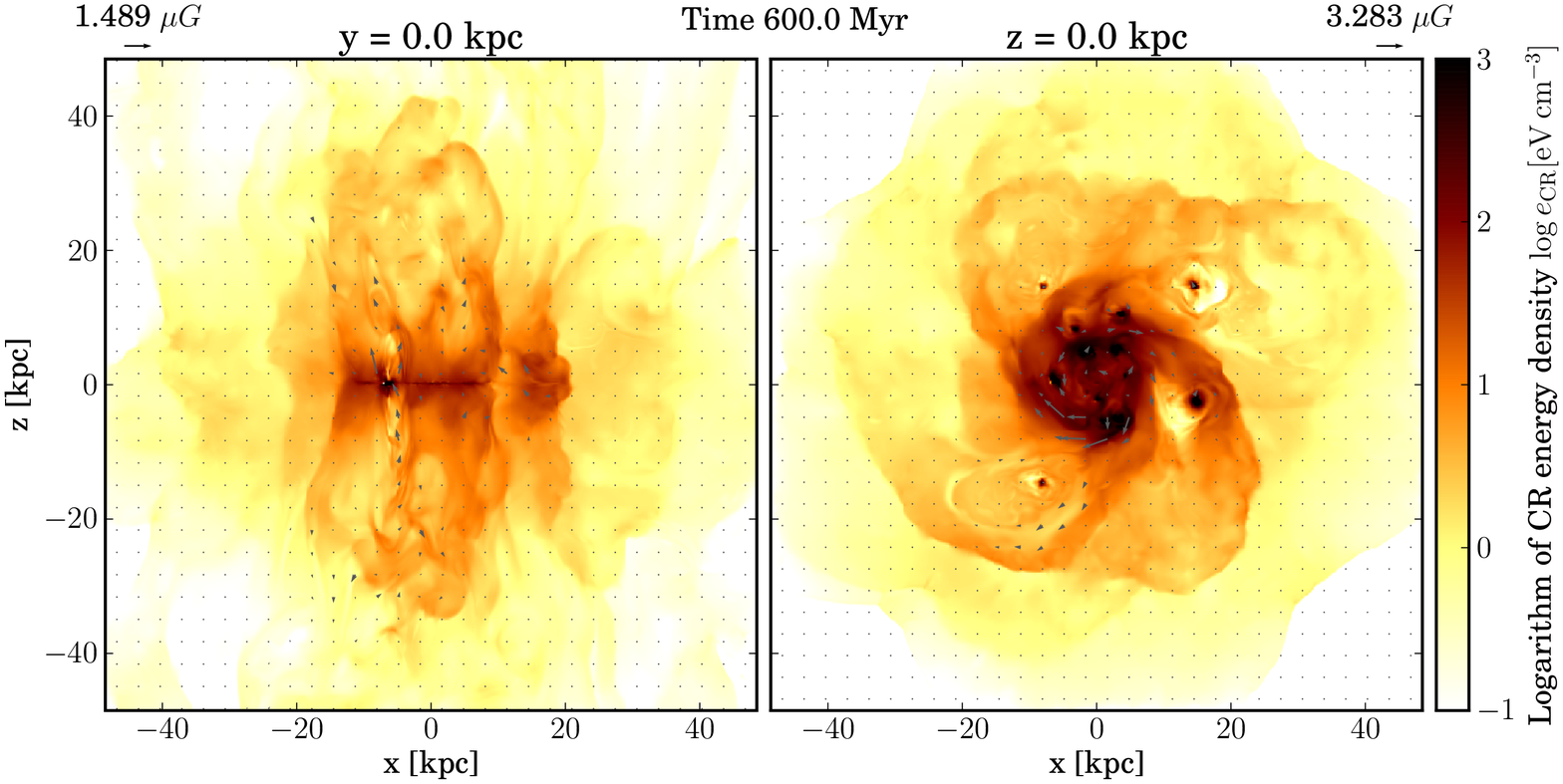}
   \caption{Vertical (left column) and horizontal  slices (right column)
      through the disk volume.  Upper panels: Logarithm of gas density and
      velocity vectors at $t = 600\Myr$. Dense gas blobs hosting star
      formation regions are apparent at the horizontal slice through the disk.
       Lower panels: Logarithm of CR energy density.  The high
       concentration of CRs at the horizontal plane coincides with the
       star forming clouds. 
   \hfill}
   \label{fig:den_cr}
\end{figure}
%
\begin{figure}[ht]
   \centering
   \includegraphics[width=0.85\columnwidth]{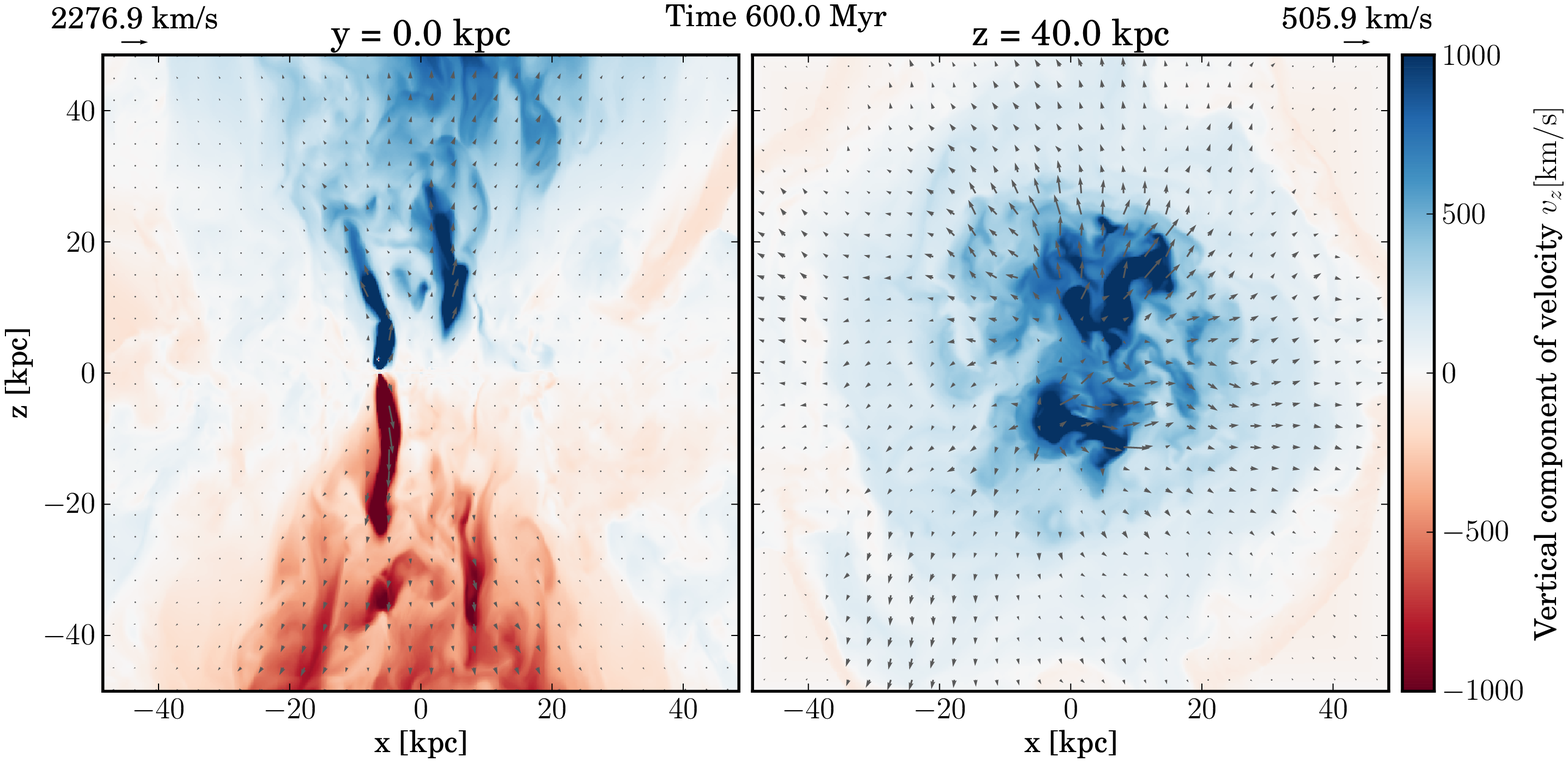}
   \includegraphics[width=0.85\columnwidth]{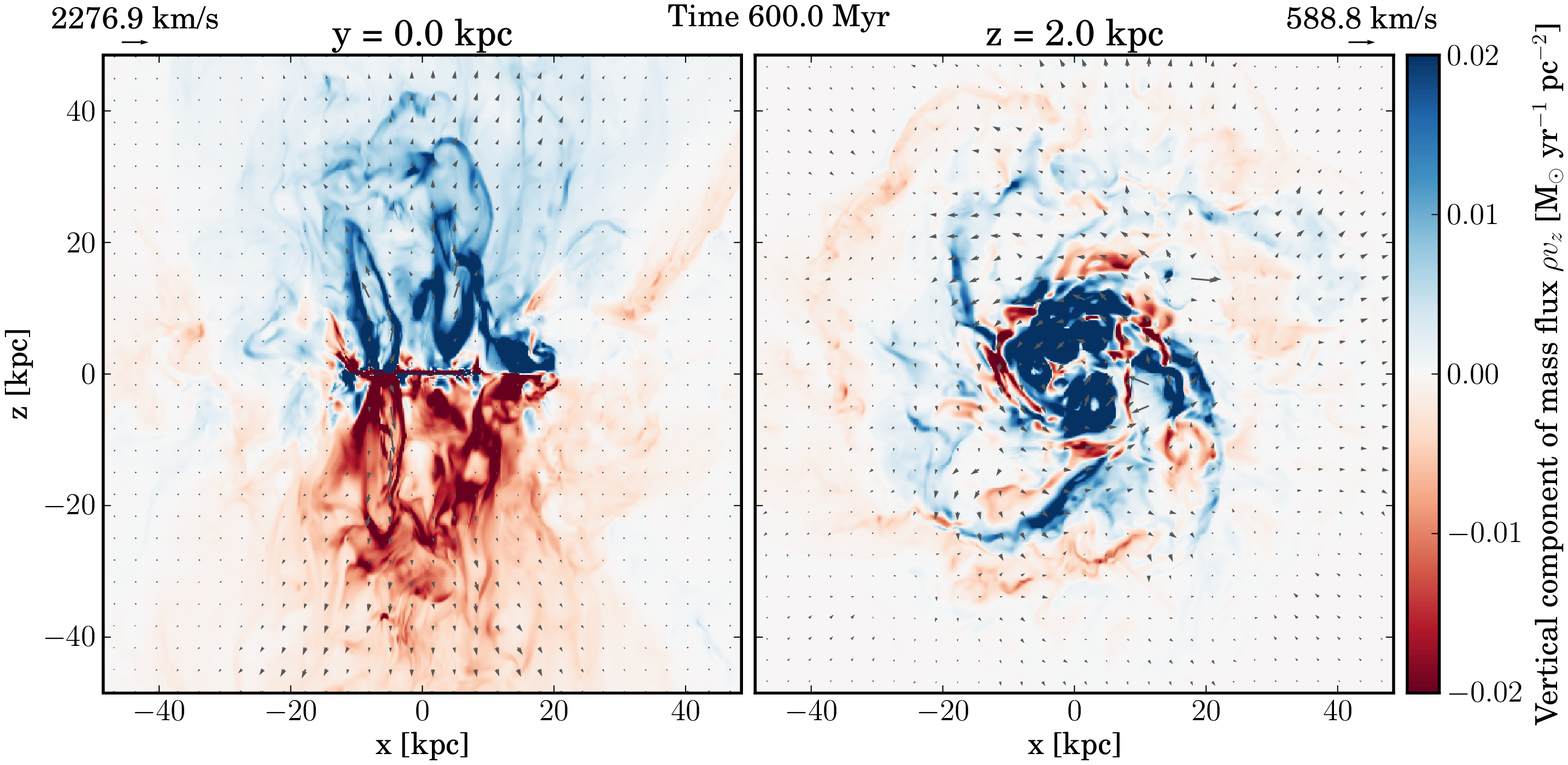}
   \includegraphics[width=0.85\columnwidth]{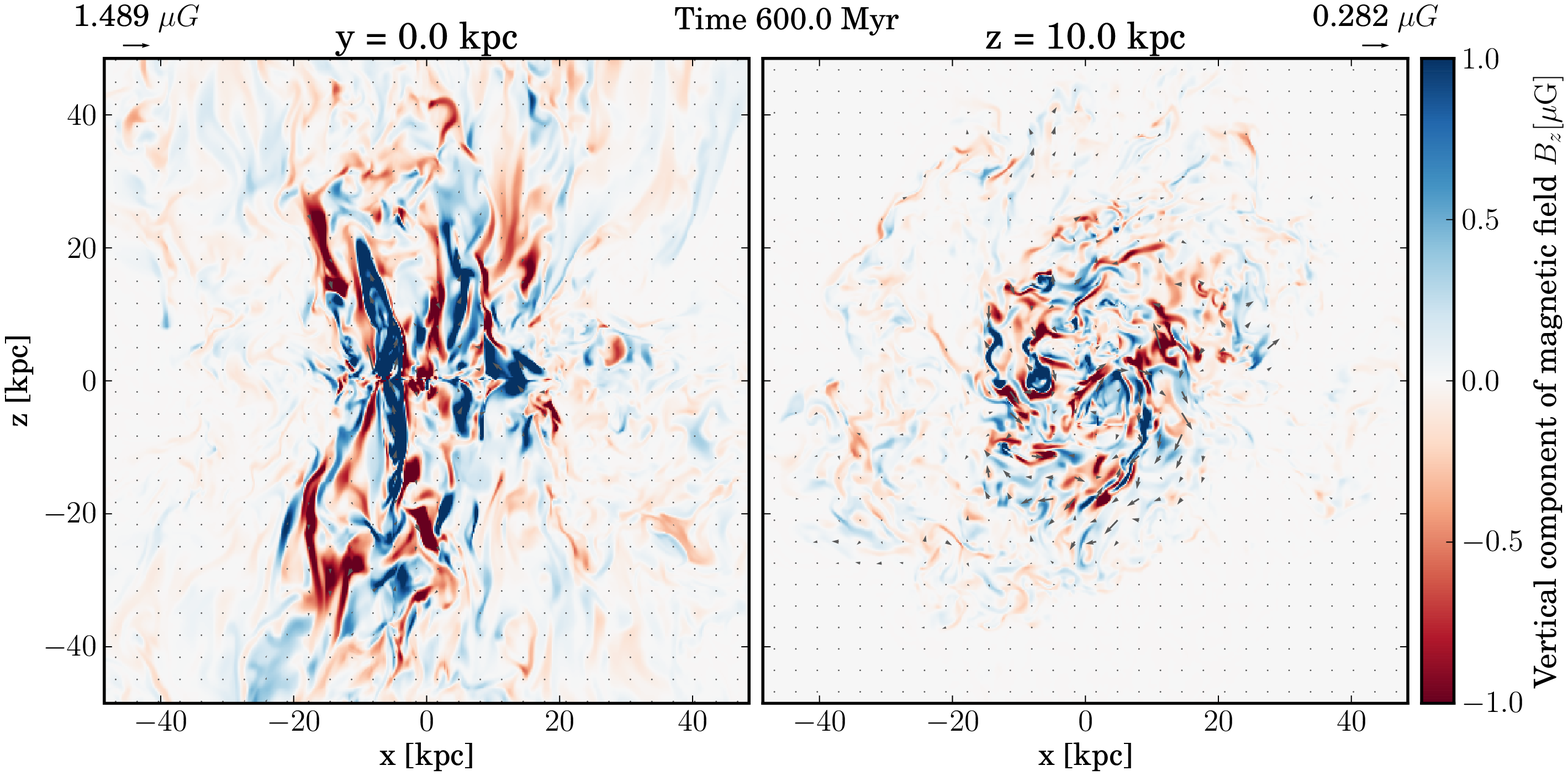}
   \caption{Vertical (left column) and horizontal (right column) maps
      at different vertical heights of wind related quantities.     
      Upper panels:  Vertical component of the velocity. Narrow streams
      of high velocity rarefied gas extend several 10 $\kpc$ above and
      below the disk. The relation of high  velocity streams to the
      underlying star formation regions is apparent. Middle panels:
      Vertical mass flux $f_{z} = \rho v_z$. Regions of high mass flux
      coincide with the  highest concentration of CRs shown in
      Fig.~\ref{fig:den_cr}. Bottom panels: Magnitude of magnetic
      field $\mathbf{B}$. Vertical filaments of $\sim1 \muG$ magnetic
      field extend to vertical distances of several tens of $\kpc$
      from the galactic plane. 
   \hfill}
   \label{fig:vz_mz_bz}
\end{figure}
%
\begin{figure}[ht]
   \centering
   \includegraphics[width=0.7\columnwidth]{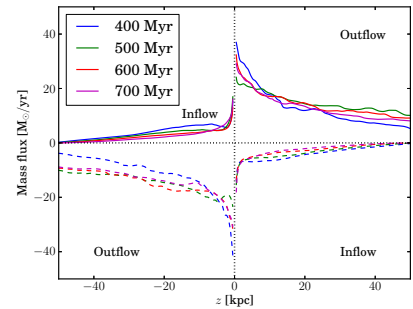}
   \caption{Horizontally integrated mass flux vs. vertical coordinate
     $z$. Solid lines denotes flux of gas moving in positive
     $z$-direction, and dashed lines denotes gas moving in negative
     $z$-direction.} 
   \label{fig:mflux}
\end{figure}
\begin{figure}[ht]
   \centering
   \includegraphics[width=0.7\columnwidth]{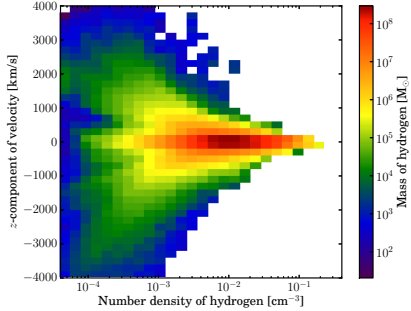}
   \caption{2D histogram of the wind mass as a function of density and
     velocity above and below the disk mid-plane. The diagram is
     constructed for two cylinders of $R=20\kpc$ located above and below the disk. The
            upper cylinder extends between $z=1\kpc$ and $z=5\kpc$, The lower
            cylinder extends between $z=-1\kpc$ and $z=-5\kpc$. It is apparent
            that high velocity gas, exceeding the escape speed ($\simeq 500
            \kms$), has densities $\leq 0.003$ H  atoms per $\cm^3$. }
\label{fig:vd_histogram}
\end{figure}
\begin{figure}[ht]
   \centering
   \includegraphics[width=0.7\columnwidth]{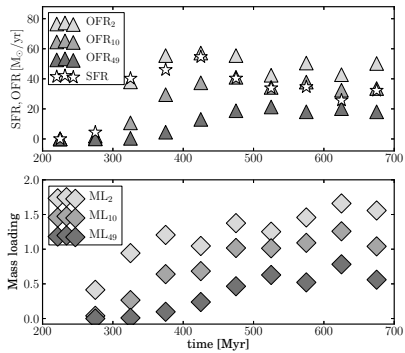}
   \caption{Upper panel: Star formation rate and mass outflow rate
     (above and below the disk) at three different altitudes of $z= \pm 2
     \kpc$, $\pm 10\kpc$, and  $z=\pm 49\kpc$ and binned in time
     intervals of $50\Myr$. Lower panel: The corresponding mass
     loading factors of order unity indicate significant outflow from
     the disk into the galactic halo. 
   }
\label{fig:moots}
\end{figure}

Initially the gaseous disk collects gas at the presumed global infall rate
$\dot{M}_{\textrm{in}}$ until it becomes locally gravitationally unstable.
Supernovae start to explode and deposit CRs in the ISM after the gas density
exceeds the critical value. After about $t\simeq 300\Myr $ the disk reaches an
equilibrium state with a star formation rate at a level of $\mathrm{SFR} \simeq
40\Msun \Myr^{-1}$. A typical snapshot of the system after $600\Myr$ of
evolution is shown in Fig.~\ref{fig:den_cr}. Most of the supernovae activity is
confined to isolated regions in \kpc-sized dense gas clouds (upper right
panel).  These regions can be also identified  as  spots of high  CR energy
density  apparent as dark brown and black patches in the face-on map (lower
right panel of  Fig.~\ref{fig:den_cr}).  One can identify about $10-12$ discrete
star formation regions with CR energy densities exceeding $\simeq100 \eV
\cm^{-3}$  dropping to  $1 \eV \cm^{-3}$ at larger distances away from the disk
(lower panels of Fig.~\ref{fig:den_cr}).  The distribution of the CR energy
density  in the galactic halo is highly non-uniform. Sharp edges of
CR-populated regions can be identified with similar structures  in the maps of
vertical mass flux and vertical magnetic field component shown in
Fig.~\ref{fig:vz_mz_bz}. 
 
The vertical streams of rarefied gas visible in gas density distribution (upper
panels of  Figs.~\ref{fig:den_cr} and~\ref{fig:vz_mz_bz}) are  accelerated, by
CRs, to high velocities  (several $10^3 \kms$). The streams can extend several
tens of $\kpc$ above and below the disk plane and significant fraction of the
outflowing gas has velocities above escape velocity and will be able to leave
the galaxy altogether. 

Maps of the mass flux $f_{z} = \rho v_z$ (mid panels  of
Fig.~\ref{fig:vz_mz_bz}) show the bimodal nature of the outflow
perpendicular to the disk plane with peak values up to $0.2 \Msun
\yr^{-1} \kpc^{-2}$ (the color scale of the mass flux panels is saturated
at only $0.02 \Msun \yr^{-1} \kpc^{-2}$ to show the wind structure
far from the disk plane). 

Streams of gas emanating from a single star forming region have a
large cross-section, visible at the horizontal slice of  $\rho v_z$ at
$z=2\kpc$. Individual SF regions generate outflows of  $5\Msun/\yr$
on average and form streams of about $3\kpc$ in radius on both sides
of the disk. In some cases the streams from neighboring SF
regions merge. The horizontal area of a single stream is a
few $\sim 10 \kpc^2$. This area multiplied by $0.2 \Msun \yr^{-1}
\kpc^{-2}$ gives a number consistent with $\sim 5\Msun/\yr$ through
the stream cross-section area, even though the average flux at
$z=2\kpc$ may be half of the peak value $\simeq 0.1 \Msun \yr^{-1}
\kpc^{-2}$ for two outflows on both sides of the disk. The mass loaded
wind is enriched with fresh CRs and is highly magnetized.  Field
strengths exceeding  $1 \muG$ can be reached naturally at distances of
$20-50\kpc$ away from the central galaxy.  Magnetic flux tubes
coherent over such large distances would be directly detectable with 
Faraday rotation measurement techniques. 

To quantify the vertical structure of the CR-driven wind we plot
in Fig.~\ref{fig:mflux} the total vertical mass flux as a function of
distance from the galactic disk from $t=400 \Myr$ to $t=700\Myr$. In
general the flow patterns are symmetric and outflow dominates inflow
by an order of magnitude. Total outflow rates decline from $\sim \pm
60 \Msun/\yr$ close to the disk plane to $ <  20 \Msun/\yr$ at $50
\kpc$.  A more detailed inspection of wind density-velocity
distribution  (Fig.~\ref{fig:vd_histogram}) shows that the density of
the CR driven wind gas is typically below $0.1 \cm^{-1}$ and the
high-velocity gas  ($v_z \geq 500   \kms$) escapes at densities $\leq
0.003$ H  atoms  $\cm^{-3}$.

In Fig.~\ref{fig:moots} we show the evolution of the star formation
rate and the integrated outflow rates at different altitudes at $z= 2
\kpc$, $10\kpc$, and $49 \kpc$ above the disk plane. After about 400
Myr the star formation rate settles to a value of
$\dot{M}_\mathrm{SFR} \sim 40 \Msun \yr^{-1}$. The mean surface
density of disk gas  within the radius 10 kpc reaches an equilibrium
value of $\sim 100 \Msun \pc^{-2}$  and SFR surface density is $ \sim 
10^{-1} \Msun \yr^{-1}  \kpc^{-2} $. These values are in good agreement
with nearby highly star forming galaxies and typical massive
high-redshift disks
\citep{1998ApJ...498..541K,2013ApJ...768...74T}. At about $10 \kpc$ 
away from the disk plane the mass outflow rate matches the star
formation rate and the galaxy average mass-loading $\eta =
\dot{M}_{\mathrm{out}}/{\rm SFR}$ becomes of order unity. Closer to
the disk plane the mass-loading is higher $\eta \sim 1.5$ and further
away from the disk plane it is still significant. It only decreases to
values of $\eta \sim 0.5$.

\section{Discussion and conclusions} \label{discussion}

We have demonstrated as a proof of principle that the injection of
only $10\% $ of SN energy in the form of CRs  and neglecting the thermal
and kinetic energy input is sufficient to drive a large scale
galactic wind in a gas-rich and highly star forming disk with
properties similar to typical star forming high-redshift galaxies. The
additional pressure gradient of the relativistic fluid - which in
contrast to heated dense gas cannot easily dissipate its energy away -
drives the formation of a strong bi-polar galactic wind with
velocities exceeding $10^3 \kms$. CRs can easily escape far from dense
regions with almost negligible energy losses and deposit their energy
and momentum in rarefied medium. This process is supported by the
CR driven break-out of field lines whose vertically
open structure is maintained by the wind. Cosmic rays  
can rapidly diffuse along these field lines far into the galactic halo. 

To check consequences of our assumption of constant diffusion
coefficients we examined the ratio of CR diffusion speed to the
Alfv\'en speed over the simulation volume. As shown in
Fig.\ref{fig:den_cr} the CR energy distribution is very smooth 
everywhere except the close to the SF regions and some apparent
shock-like structures, especially when compared to the gas density
distribution. Vertical stratification of the CR energy distribution
can be observed only far from star formation regions and the
corresponding scale height is of the order of a few kpc.  
The smooth distribution of CRs is a consequence of a highly irregular  
vertical magnetic field  component, guiding the field aligned CR
diffusion in the vertical direction. We have found that
$v_\mathrm{diff}/v_\mathrm{A} \leq 1 $  in a major part of the
simulation volume. Consequently, $v_\mathrm{diff}/v_\mathrm{A} > 1 $
in regions of galactic disk with high gas density. As we mentioned
already, we consider the limitation of the diffusion velocity by Alfv\'en
waves less restrictive in the disk volume, because neutrals  are
supposed to reduce the level of CR self-excited Alfv\'enic turbulence.  
Moreover, we note that steep gradients of CR energy density  around
star formation regions provide a specific feedback mechanism. High
production rates of CRs implies fast expansion of overpressured
regions, and subsequently enhanced CR advective expansion leads to
a reduction of the CR energy density gradient. The fast expansion of
CR overpressured bubbles increases the Alfv\'en speed locally, leading
to a reduction of the ratio $v_\mathrm{diff}/v_\mathrm{A}$. 

The  CR driving is so significant that the mass outflow rate can
become of the same order as the star formation rate in the galactic
disk, even in our simplified setup where the disk plane is more or
less treated as an inner boundary condition and thermal as well as
kinetic feedback from stellar evolution and supernovae have been
neglected entirely. Based on our simulations we can conclude that
relativistic particles accelerated in supernova remnants in
combination with a strong magnetic fields (typical for high-redshift
galaxies \citep{2008Natur.454..302B}) provide a natural  and efficient
mechanism to help explaining the ubiquitously observed mass-loaded
galactic winds in high-redshift galaxies (e.g.
\citep{2003ApJ...588...65S,2012ApJ...752..111N}) as well as the highly 
magnetized medium surrounding these galaxies
\citep{2013arXiv1307.2250B}. The efficiency - in terms of mass loading
- of this wind driving process appears to be comparable to momentum
and energy driving from stellar evolution and supernovae explosions 
(e.g. \citealp{2012MNRAS.421.3522H,2012MNRAS.426..140D,2013ApJ...770...25A})
and requires further investigation.

\begin{acknowledgements}
We thank the anonymous referee for valuable comments on the
manuscript. TN acknowledges support from the DFG priority program
SPP1653. MH acknowledges the generous support of Alexander von
Humboldt Foundation and University Observatory, Ludwigs-Maximilian
University for kind hospitality. All of the post-processing and
visualization was carried out using the data analysis and
visualization package \yt{}\footnote{\url{http://yt-project.org/}}
by~\cite{yt}.This work was partially supported by Polish Ministry of
Science and Higher Education through the grant N203 511038. This work
is also a part of POWIEW project supported by the European Regional
Development Fund in the Innovative Economy Programme
(POIG.02.03.00-00-018/08).  This research was supported in part by
PL-Grid Infrastructure. 
\end{acknowledgements}

\bibliographystyle{apj}
 

\begin{thebibliography}{50}
\expandafter\ifx\csname natexlab\endcsname\relax\def\natexlab#1{#1}\fi

\bibitem[{{Agertz} {et~al.}(2013){Agertz}, {Kravtsov}, {Leitner}, \&
  {Gnedin}}]{2013ApJ...770...25A}
{Agertz}, O., {Kravtsov}, A.~V., {Leitner}, S.~N., \& {Gnedin}, N.~Y. 2013,
  \apj, 770, 25

\bibitem[{{Allen} \& {Santillan}(1991)}]{1991RMxAA..22..255A}
{Allen}, C., \& {Santillan}, A. 1991, Revista Mexicana de Astronomia y
  Astrofisica, 22, 255

\bibitem[{{Behroozi} {et~al.}(2010){Behroozi}, {Conroy}, \&
  {Wechsler}}]{2010ApJ...717..379B}
{Behroozi}, P.~S., {Conroy}, C., \& {Wechsler}, R.~H. 2010, \apj, 717, 379

\bibitem[{{Bell}(1978)}]{1978MNRAS.182..147B}
{Bell}, A.~R. 1978, \mnras, 182, 147

\bibitem[{{Berezinskii} {et~al.}(1990){Berezinskii}, {Bulanov}, {Dogiel}, \&
  {Ptuskin}}]{1990acr..book.....B}
{Berezinskii}, V.~S., {Bulanov}, S.~V., {Dogiel}, V.~A., \& {Ptuskin}, V.~S.
  1990, {Astrophysics of cosmic rays} (Amsterdam: North-Holland, 1990, edited
  by Ginzburg, V.L.)

\bibitem[{{Bernet} {et~al.}(2013){Bernet}, {Miniati}, \&
  {Lilly}}]{2013arXiv1307.2250B}
{Bernet}, M.~L., {Miniati}, F., \& {Lilly}, S.~J. 2013, ArXiv e-prints

\bibitem[{{Bernet} {et~al.}(2008){Bernet}, {Miniati}, {Lilly}, {Kronberg}, \&
  {Dessauges-Zavadsky}}]{2008Natur.454..302B}
{Bernet}, M.~L., {Miniati}, F., {Lilly}, S.~J., {Kronberg}, P.~P., \&
  {Dessauges-Zavadsky}, M. 2008, \nat, 454, 302

\bibitem[{{Blandford} \& {Ostriker}(1978)}]{1978ApJ...221L..29B}
{Blandford}, R.~D., \& {Ostriker}, J.~P. 1978, \apjl, 221, L29

\bibitem[{{Breitschwerdt} {et~al.}(2002){Breitschwerdt}, {Dogiel}, \&
  {V{\"o}lk}}]{2002A&A...385..216B}
{Breitschwerdt}, D., {Dogiel}, V.~A., \& {V{\"o}lk}, H.~J. 2002, \aap, 385, 216

\bibitem[{{Breitschwerdt} {et~al.}(1991){Breitschwerdt}, {McKenzie}, \&
  {Voelk}}]{1991A&A...245...79B}
{Breitschwerdt}, D., {McKenzie}, J.~F., \& {Voelk}, H.~J. 1991, \aap, 245, 79

\bibitem[{{Brook} {et~al.}(2011){Brook}, {Governato}, {Ro{\v s}kar}, {Stinson},
  {Brooks}, {Wadsley}, {Quinn}, {Gibson}, {Snaith}, {Pilkington}, {House}, \&
  {Pontzen}}]{2011MNRAS.415.1051B}
{Brook}, C.~B., {et~al.} 2011, \mnras, 415, 1051

\bibitem[{{Croton} {et~al.}(2006){Croton}, {Springel}, {White}, {De Lucia},
  {Frenk}, {Gao}, {Jenkins}, {Kauffmann}, {Navarro}, \&
  {Yoshida}}]{2006MNRAS.365...11C}
{Croton}, D.~J., {et~al.} 2006, \mnras, 365, 11

\bibitem[{{Dalla Vecchia} \& {Schaye}(2012)}]{2012MNRAS.426..140D}
{Dalla Vecchia}, C., \& {Schaye}, J. 2012, \mnras, 426, 140

\bibitem[{{Dekel} \& {Silk}(1986)}]{1986ApJ...303...39D}
{Dekel}, A., \& {Silk}, J. 1986, \apj, 303, 39

\bibitem[{{Dorfi} \& {Breitschwerdt}(2012)}]{2012A&A...540A..77D}
{Dorfi}, E.~A., \& {Breitschwerdt}, D. 2012, \aap, 540, A77

\bibitem[{{Evans} \& {Hawley}(1988)}]{1988ApJ...332..659E}
{Evans}, C.~R., \& {Hawley}, J.~F. 1988, \apj, 332, 659

\bibitem[{{Everett} {et~al.}(2008){Everett}, {Zweibel}, {Benjamin}, {McCammon},
  {Rocks}, \& {Gallagher}}]{2008ApJ...674..258E}
{Everett}, J.~E., {Zweibel}, E.~G., {Benjamin}, R.~A., {McCammon}, D., {Rocks},
  L., \& {Gallagher}, III, J.~S. 2008, \apj, 674, 258

\bibitem[{{Ferriere}(1998)}]{1998ApJ...497..759F}
{Ferriere}, K. 1998, \apj, 497, 759

\bibitem[{{Genzel} {et~al.}(2011){Genzel}, {Newman}, {Jones}, {F{\"o}rster
  Schreiber}, {Shapiro}, {Genel}, {Lilly}, {Renzini}, {Tacconi}, {Bouch{\'e}},
  {Burkert}, {Cresci}, {Buschkamp}, {Carollo}, {Ceverino}, {Davies}, {Dekel},
  {Eisenhauer}, {Hicks}, {Kurk}, {Lutz}, {Mancini}, {Naab}, {Peng},
  {Sternberg}, {Vergani}, \& {Zamorani}}]{2011ApJ...733..101G}
{Genzel}, R., {et~al.} 2011, \apj, 733, 101

\bibitem[{{Guo} {et~al.}(2010){Guo}, {White}, {Li}, \&
  {Boylan-Kolchin}}]{2010MNRAS.404.1111G}
{Guo}, Q., {White}, S., {Li}, C., \& {Boylan-Kolchin}, M. 2010, \mnras, 404,
  1111

\bibitem[{{Hanasz} {et~al.}(2004){Hanasz}, {Kowal}, {Otmianowska-Mazur}, \&
  {Lesch}}]{2004ApJ...605L..33H}
{Hanasz}, M., {Kowal}, G., {Otmianowska-Mazur}, K., \& {Lesch}, H. 2004, \apjl,
  605, L33

\bibitem[{{Hanasz} {et~al.}(2010){Hanasz}, {Kowalik}, {W{\'o}lta{\'n}ski}, \&
  {Paw{\l}aszek}}]{piernik1}
{Hanasz}, M., {Kowalik}, K., {W{\'o}lta{\'n}ski}, D., \& {Paw{\l}aszek}, R.
  2010, in EAS Publications Series, Vol.~42, EAS Publications Series, ed.
  K.~{Go{\'z}dziewski}, A.~{Niedzielski}, \& J.~{Schneider}, 275--280

\bibitem[{{Hanasz} \& {Lesch}(2003)}]{2003A&A...412..331H}
{Hanasz}, M., \& {Lesch}, H. 2003, \aap, 412, 331

\bibitem[{{Hanasz} {et~al.}(2009){Hanasz}, {W{\'o}lta{\'n}ski}, \&
  {Kowalik}}]{2009ApJ...706L.155H}
{Hanasz}, M., {W{\'o}lta{\'n}ski}, D., \& {Kowalik}, K. 2009, \apjl, 706, L155

\bibitem[{{Hopkins} \& {Beacom}(2006)}]{2006ApJ...651..142H}
{Hopkins}, A.~M., \& {Beacom}, J.~F. 2006, \apj, 651, 142

\bibitem[{{Hopkins} {et~al.}(2012){Hopkins}, {Quataert}, \&
  {Murray}}]{2012MNRAS.421.3522H}
{Hopkins}, P.~F., {Quataert}, E., \& {Murray}, N. 2012, \mnras, 421, 3522

\bibitem[{Huang \& Greengard(1999)}]{doi:10.1137/S1064827598346235}
Huang, J., \& Greengard, L. 1999, SIAM Journal on Scientific Computing, 21,
  1551

\bibitem[{{James}(1977)}]{1977JCoPh..25...71J}
{James}, R.~A. 1977, Journal of Computational Physics, 25, 71

\bibitem[{{Jin} \& {Xin}(1995)}]{jin-xin-95}
{Jin}, S., \& {Xin}, Z. 1995, Comm. Pure Appl. Math., 48, 235

\bibitem[{{Kennicutt}(1998)}]{1998ApJ...498..541K}
{Kennicutt}, R.~C. 1998, \apj, 498, 541

\bibitem[{{Krymskii}(1977)}]{1977DoSSR.234.1306K}
{Krymskii}, G.~F. 1977, Akademiia Nauk SSSR Doklady, 234, 1306

\bibitem[{{Kulpa-Dybe{\l}} {et~al.}(2011){Kulpa-Dybe{\l}}, {Otmianowska-Mazur},
  {Kulesza-{\.Z}ydzik}, {Hanasz}, {Kowal}, {W{\'o}lta{\'n}ski}, \&
  {Kowalik}}]{2011ApJ...733L..18K}
{Kulpa-Dybe{\l}}, K., {Otmianowska-Mazur}, K., {Kulesza-{\.Z}ydzik}, B.,
  {Hanasz}, M., {Kowal}, G., {W{\'o}lta{\'n}ski}, D., \& {Kowalik}, K. 2011,
  \apjl, 733, L18

\bibitem[{{Kulsrud} \& {Pearce}(1969)}]{1969ApJ...156..445K}
{Kulsrud}, R., \& {Pearce}, W.~P. 1969, \apj, 156, 445

\bibitem[{{Larson}(1974)}]{1974MNRAS.169..229L}
{Larson}, R.~B. 1974, \mnras, 169, 229

\bibitem[{{Moster} {et~al.}(2013){Moster}, {Naab}, \&
  {White}}]{2013MNRAS.428.3121M}
{Moster}, B.~P., {Naab}, T., \& {White}, S.~D.~M. 2013, \mnras, 428, 3121

\bibitem[{{Moster} {et~al.}(2010){Moster}, {Somerville}, {Maulbetsch}, {van den
  Bosch}, {Macci{\`o}}, {Naab}, \& {Oser}}]{2010ApJ...710..903M}
{Moster}, B.~P., {Somerville}, R.~S., {Maulbetsch}, C., {van den Bosch}, F.~C.,
  {Macci{\`o}}, A.~V., {Naab}, T., \& {Oser}, L. 2010, \apj, 710, 903

\bibitem[{{Murray} {et~al.}(2005){Murray}, {Quataert}, \&
  {Thompson}}]{2005ApJ...618..569M}
{Murray}, N., {Quataert}, E., \& {Thompson}, T.~A. 2005, \apj, 618, 569

\bibitem[{{Newman} {et~al.}(2012{\natexlab{a}}){Newman}, {Shapiro Griffin},
  {Genzel}, {Davies}, {F{\"o}rster-Schreiber}, {Tacconi}, {Kurk}, {Wuyts},
  {Genel}, {Lilly}, {Renzini}, {Bouch{\'e}}, {Burkert}, {Cresci}, {Buschkamp},
  {Carollo}, {Eisenhauer}, {Hicks}, {Lutz}, {Mancini}, {Naab}, {Peng}, \&
  {Vergani}}]{2012ApJ...752..111N}
{Newman}, S.~F., {et~al.} 2012{\natexlab{a}}, \apj, 752, 111

\bibitem[{{Newman} {et~al.}(2012{\natexlab{b}}){Newman}, {Genzel},
  {F{\"o}rster-Schreiber}, {Shapiro Griffin}, {Mancini}, {Lilly}, {Renzini},
  {Bouch{\'e}}, {Burkert}, {Buschkamp}, {Carollo}, {Cresci}, {Davies},
  {Eisenhauer}, {Genel}, {Hicks}, {Kurk}, {Lutz}, {Naab}, {Peng}, {Sternberg},
  {Tacconi}, {Vergani}, {Wuyts}, \& {Zamorani}}]{2012ApJ...761...43N}
---. 2012{\natexlab{b}}, \apj, 761, 43

\bibitem[{{Parker}(1992)}]{1992ApJ...401..137P}
{Parker}, E.~N. 1992, \apj, 401, 137

\bibitem[{{Pen} {et~al.}(2003){Pen}, {Arras}, \& {Wong}}]{2003ApJS..149..447P}
{Pen}, U.-L., {Arras}, P., \& {Wong}, S. 2003, \apjs, 149, 447

\bibitem[{{Pettini} {et~al.}(2000){Pettini}, {Steidel}, {Adelberger},
  {Dickinson}, \& {Giavalisco}}]{2000ApJ...528...96P}
{Pettini}, M., {Steidel}, C.~C., {Adelberger}, K.~L., {Dickinson}, M., \&
  {Giavalisco}, M. 2000, \apj, 528, 96

\bibitem[{{Shapley} {et~al.}(2003){Shapley}, {Steidel}, {Pettini}, \&
  {Adelberger}}]{2003ApJ...588...65S}
{Shapley}, A.~E., {Steidel}, C.~C., {Pettini}, M., \& {Adelberger}, K.~L. 2003,
  \apj, 588, 65

\bibitem[{{Siejkowski} {et~al.}(2010){Siejkowski}, {Soida},
  {Otmianowska-Mazur}, {Hanasz}, \& {Bomans}}]{2010A&A...510A..97S}
{Siejkowski}, H., {Soida}, M., {Otmianowska-Mazur}, K., {Hanasz}, M., \&
  {Bomans}, D.~J. 2010, \aap, 510, A97

\bibitem[{{Steidel} {et~al.}(2010){Steidel}, {Erb}, {Shapley}, {Pettini},
  {Reddy}, {Bogosavljevi{\'c}}, {Rudie}, \& {Rakic}}]{2010ApJ...717..289S}
{Steidel}, C.~C., {Erb}, D.~K., {Shapley}, A.~E., {Pettini}, M., {Reddy}, N.,
  {Bogosavljevi{\'c}}, M., {Rudie}, G.~C., \& {Rakic}, O. 2010, \apj, 717, 289

\bibitem[{{Strong} {et~al.}(2007){Strong}, {Moskalenko}, \&
  {Ptuskin}}]{2007ARNPS..57..285S}
{Strong}, A.~W., {Moskalenko}, I.~V., \& {Ptuskin}, V.~S. 2007, Annual Review
  of Nuclear and Particle Science, 57, 285

\bibitem[{{Tacconi} {et~al.}(2013){Tacconi}, {Neri}, {Genzel}, {Combes},
  {Bolatto}, {Cooper}, {Wuyts}, {Bournaud}, {Burkert}, {Comerford}, {Cox},
  {Davis}, {F{\"o}rster Schreiber}, {Garc{\'{\i}}a-Burillo}, {Gracia-Carpio},
  {Lutz}, {Naab}, {Newman}, {Omont}, {Saintonge}, {Shapiro Griffin}, {Shapley},
  {Sternberg}, \& {Weiner}}]{2013ApJ...768...74T}
{Tacconi}, L.~J., {et~al.} 2013, \apj, 768, 74

\bibitem[{{Turk} {et~al.}(2011){Turk}, {Smith}, {Oishi}, {Skory}, {Skillman},
  {Abel}, \& {Norman}}]{yt}
{Turk}, M.~J., {Smith}, B.~D., {Oishi}, J.~S., {Skory}, S., {Skillman}, S.~W.,
  {Abel}, T., \& {Norman}, M.~L. 2011, \apjs, 192, 9

\bibitem[{{Uhlig} {et~al.}(2012){Uhlig}, {Pfrommer}, {Sharma}, {Nath},
  {En{\ss}lin}, \& {Springel}}]{2012MNRAS.423.2374U}
{Uhlig}, M., {Pfrommer}, C., {Sharma}, M., {Nath}, B.~B., {En{\ss}lin}, T.~A.,
  \& {Springel}, V. 2012, \mnras, 423, 2374

\bibitem[{{Weiner} {et~al.}(2009){Weiner}, {Coil}, {Prochaska}, {Newman},
  {Cooper}, {Bundy}, {Conselice}, {Dutton}, {Faber}, {Koo}, {Lotz}, {Rieke}, \&
  {Rubin}}]{2009ApJ...692..187W}
{Weiner}, B.~J., {et~al.} 2009, \apj, 692, 187

\end{thebibliography}

\end{document}